# Nanoscale

## ARTICLE

# Spatially uniform resistance switching of low current, high endurance titanium-niobium-oxide memristors

Suhas Kumar,[1] Noraica Davila,[1] Ziwen Wang,[2] Xiaopeng Huang,[1] John Paul Strachan,[1] David Vine,[3] A. L. David Kilcoyne,[3] Yoshio Nishi[2] and R. Stanley Williams[1]

We analyzed micrometer-scale titanium-niobium-oxide prototype memristors, which exhibited low write-power (<3 μW) and energy (<200 fJ/bit/μm$^2$), low read-power (~nW), and high endurance (>millions of cycles). To understand their physico-chemical operating mechanisms, we performed *in-operando* synchrotron x-ray transmission nanoscale spectromicroscopy using an ultra-sensitive time-multiplexed technique. We observed only spatially uniform material changes during cell operation, in sharp contrast to the frequently detected formation of a localized conduction channel in transition-metal-oxide memristors. We also associated the response of assigned spectral features distinctly to non-volatile storage (resistance change) and writing of information (application of voltage and Joule heating). These results provide critical insights into high-performance memristors that will aid in device design, scaling and predictive circuit-modeling, all of which are essential for the widespread deployment of successful memristor applications.

## Introduction and electrical characterization

The Memristors are frontrunners for upcoming-generation storage-class memory.[1,2] Cell-level integration of non-volatile storage, current limiting capability and tunable electrical isolation will directly address critical memory performance challenges such as operating power, data density and circuit overheads.[3-5] Integration of multiple materials and device engineering to achieve such enhanced memristor performance is an active area of research.[3,5-7] Here we fabricated crosspoint micrometer-scale memristors with thin films of $Nb_2O_5/NbO_2/TiNO_{0.5}$ as the active material stack sandwiched between two Pt electrodes, and characterized each using x-ray transmission spectral signatures from the different components (Figures 1a-1b).[8,9] The titanium oxynitride film stoichiometry was inferred from separate photoemission measurements and the niobium oxide films were also separately characterized (Supplementary Material). Niobium oxides and titanium oxynitride are of interest for their applications in non-volatile switching, limiting the current during switching and providing tunable electrical isolation through non-linearity in conduction.[10-15] The bipolar non-volatile resistance switching occurred at low power levels <3 μW between non-linear conductances in both the low resistance (ON) and high-resistance (OFF) states (Figure 1c). The non-linearity is of particular utility since cells at half-bias during writing within a crossbar have much smaller leakage currents than if the voltage characteristics of the cell under study obtained by sweeping the current. Inset is the same plot in a semi-logarithmic scale. ON and OFF states refer to the low resistance and high resistance states, respectively. Voltage references are discussed in the text.

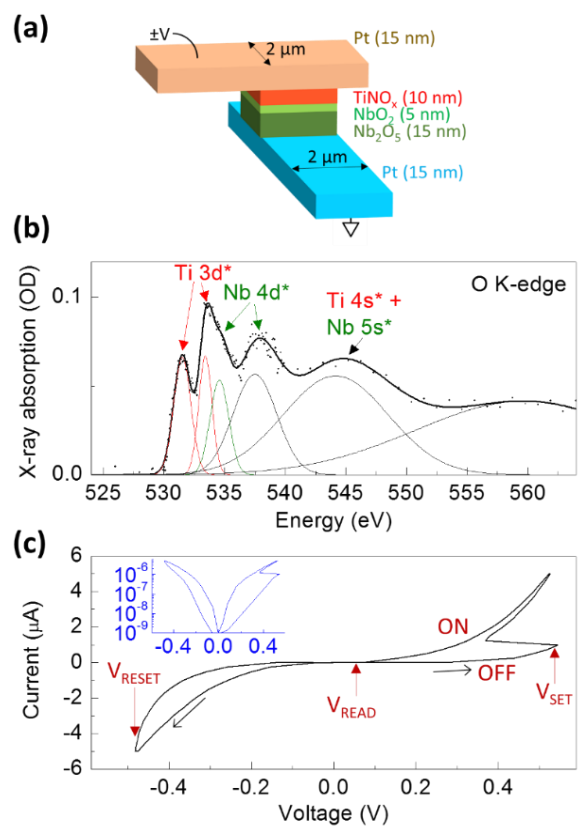

**Figure 1: Cell schematic, chemical and electrical characterization.** *(a) Schematic of the cell structure and the material layers, showing the electrical addressing scheme. (b) O K-edge absorption spectrum through the crosspoint material stack. The sub-bands are shown and if distinct, identified with the orbital transitions (from O 1s) that cause them. (c) Current-*

[1]Hewlett Packard Labs, 1501 Page Mill Rd, Palo Alto, CA 94304, USA
[2]Stanford University, Stanford, CA 94305, USA
[3]Lawrence Berkeley National Laboratory, Berkeley, CA 94720, USA
Electronic Supplementary Information (ESI) available

conductances were linear.[16] In addition, a cell for which the resistance level is 'read'/measured at a lower voltage ($V_{READ}$) dissipates much lower power (~nW) than during the SET/RESET operations (~µW). We performed high-speed switching experiments and showed SET and RESET operations in < 20 ns using < 1.5 V pulse amplitudes and an associated energy of 0.3-0.8 pJ/bit (75-200 fJ/µm²) including all parasitics (Supplementary Material).

## *In-operando* x-ray characterization technique

One of the greatest challenges in physically understanding memristor operation has been the extremely localized, low signal, atomic-scale material changes associated with large changes in the cell resistance, the result of which is uncertainty and controversy on the details of the operating mechanisms.[4,17-23] Experiments are usually performed using destructive techniques like cross sectional electron microscopy,[24] non-standard device construction,[23] and amplifying the material changes by using stronger operating conditions.[18,19] To overcome these limitations, we developed an ultra-sensitive measurement technique to probe the electronic, structural, and chemical properties during regular operation of a cell.[18] In order to achieve the necessary spatial and spectral resolution, we employed a synchrotron-based scanning transmission x-ray microscope (STXM) with spatial resolution of <30 nm and spectral resolution of ~70 meV.[25] Prototype devices for this experiment were fabricated atop 200 nm thick freely suspended $Si_3N_4$ membranes to enable x-ray transmission.[26,27] Further, in order to overcome the signal-limiting issues of spatial drift, background absorption changes, stochasticity in cell operation and drift in cell behavior, we constructed an *in-operando* time-multiplexed experimental setup (Figure 2b). We incorporated an adaptive cell-switching technique that utilized feedback-enabled resistance switching together with a verification read, while synchronously gating the detector signal (from the asynchronous synchrotron x-ray pulses) into two different counters corresponding to physical measurements only when the target resistance states were successfully achieved. The result was an integration of signal at every spatial and spectral position over many verified switching events, which averages stochastic processes during cell switching. More importantly, the time-multiplexing provided a reduction in effects due to spatial drift of the sample and other background changes by over 5 orders of magnitude, detailed elsewhere.[18] The adaptive system also corrected for temporal drifts in cell switching behavior, which is another major source of distortion in such measurements.

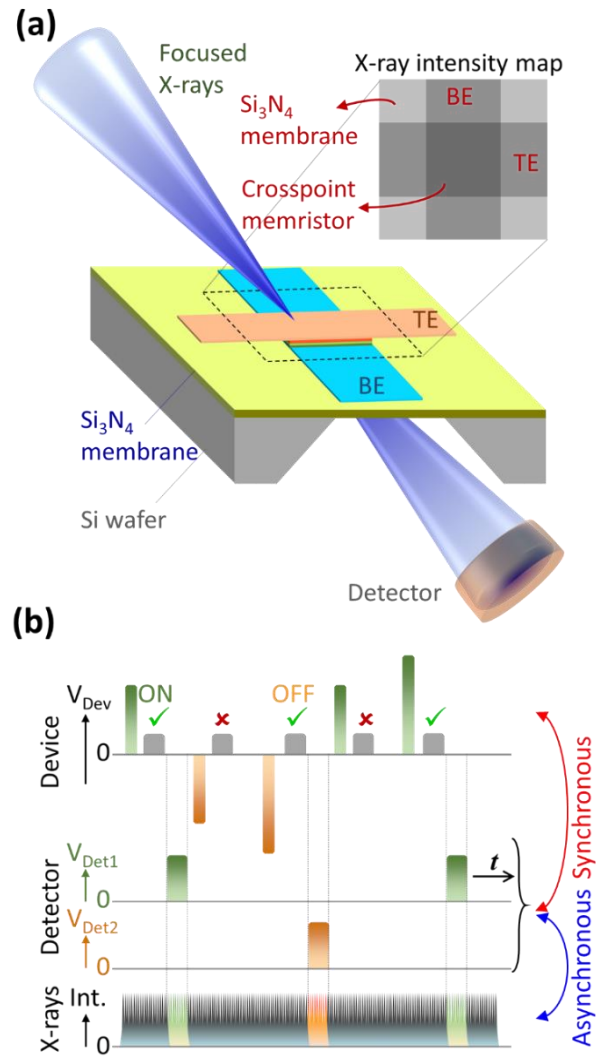

**Figure 2: STXM and time-multiplexed setup.** *(a) Schematic of the STXM measurement setup and the cell under measurement. Also shown is a schematic of the x-ray intensity map obtained by imaging the crosspoint region of cell structure, with the top electrode (TE) and the bottom electrode (BE) identified. (b) Schematic of the adaptive time-multiplexed technique applied to the STXM measurements. The device voltage ($V_{Dev}$) consisted of bipolar voltage pulses (green and orange) to continually switch the resistance of the cell, and subsequently read the resistance (grey pulses) to verify that a pre-determined threshold for the ON or OFF state of the device was reached ('✓'), while a failure ('✗') resulted in repeated attempts with larger voltage pulses. Detector voltages ($V_{Det1}$ and $V_{Det2}$) were synchronously applied only upon verification of a successful resistance change to store signals from the ON and OFF states in two separate counters, obtained by measuring the asynchronous x-ray intensity (x-rays, int.) during detection.*

## Results of x-ray characterization

**Resistance switching effect (information storage)**

Using the setup described above, we switched the resistance of the same cell 4 million times (Figure 3a) while simultaneously mapping the crosspoint in the ON and OFF states of the cell (Figure 3b), defined by resistance as $R_{ON}$<60 kΩ and $R_{OFF}$>1 MΩ, attained by applying pulses of width 5 μs and amplitude 0.4-1.4 V. The raw x-ray transmission maps do not show differences between each other observable by eye, whereas the logarithmic ratio of the two, or optical density (OD), revealed a small but detectable change in the x-ray absorption uniformly dispersed over the crosspoint region (Figures 3c). The measured OD distribution inside the crosspoint (Figure 3d) was shift to higher values, while that outside the crosspoint was centered at zero, revealing a spectral response to the change in resistance only inside the crosspoint. Both the OD distributions were essentially Gaussian with nearly identical standard deviations, indicating that the spectral response was spatially uniform (further confirmed in Figures S9-S13). The measured sample standard deviations of the OD distributions (S ≈ 0.72×10$^{-3}$) were comparable to the mean of the OD inside the crosspoint (M ≈ 0.72×10$^{-3}$), while the standard error of the mean OD was estimated from the sample standard deviation as $M_{err} = S/\sqrt{N}$ ≈ 0.01×10$^{-3}$ for N>2500 representing many measurement cycles. The change in resistance is thus characterized by a statistically significant rigid shift of the OD distribution inside the crosspoint. The O K-edge spectra of the region within the crosspoint in the ON and OFF states of the cell (Figure 3e) are very similar, while their difference displays a spectral shift at the rising edge (~530 eV) attributed to the conduction band minimum (CBM), which mainly originates from Ti-3d – O-2p bonds. This represents a small but detectable and reproducible shift of the electronic density of states in the OFF versus the ON state, with the downshift in the CBM corresponding to an increased conductivity uniformly throughout the crosspoint region. This difference is <0.5% of the background signal, which could not have been detected without employing the time-multiplexed technique in combination with the high chemical/spectral resolution of the STXM. The spatially uniform changes associated with resistance switching are a notable result, given that several simpler memristor material stacks (e.g.: Ta/TaO$_x$) have revealed localized conduction channel formation.[4,19,24] This observation highlights the importance of performing non-destructive, *in-operando* measurements on cells that undergo standard operation (without having to use exaggerated operating conditions or destructive characterization) to understand and model the cell operation. In principle, this technique can also probe the changes from bonds involving nitrogen (using the N K-edge when the device resides on a non-nitride window) and directly the oxidation states of the metal atoms (using the metal L-edges) as well.

**Joule heating and electric field effects (writing of information)**

We further studied the effect of the applied electric fields (voltage pulses) on the material stack. While the preceding experiment measured the changes to the materials only after the application of the voltage pulses and the resulting resistance change, here we recorded the device properties during the voltage pulses. Two new types of measurements were conducted. In the first set, we used 1 V pulses of only positive polarity on a device that was left in the ON state (to allow for larger current flow and easier detection) and also recorded a reference signal corresponding to no application of a voltage (Figure S7). A mapping of the effect of voltage revealed a uniform statistically significant change within the crosspoint region (Figure 4a-4b and S6), similar to that observed during the change in resistance. This indicated that current flowed uniformly through the entire crosspoint area. The changes in the spectral difference (Figure 4e, marked $S_{E+}$-$S_0$) appear more pronounced in the energies corresponding to the Nb-O bonds, possibly by the expansion of the lattice during Joule heating consistent with the temperature-driven non-linear behavior of $NbO_x$.[10] In the second set of experiments, we measured the material changes due to pulses of opposite polarity (±1 V), and we again found spatially uniform and statistically significant effects (Figures 4c-4d, S6 and S8). The changes in the spectra appear to be chemically localized to the Ti-O bonds around 530 eV (Figure 4e, marked $S_{E+}$-$S_{E-}$). This effect suggests an increase in conductivity by downshifting in energy of the conduction band upon application of a positive voltage, consistent with a SET process, while the effect is reversed during a negative voltage. This is also consistent with electric-field-driven migration of O vacancies into and out of $TiO_x$,[28] although not uniquely constrained by this data. These changes are likely dominated by electric-field-driven effects, given that the Joule heating from equal voltages of opposite polarity should be comparable (Figure 1c), and are significantly less pronounced upon taking the ratio of the signals. Also, this is evident since the Joule-heating-driven signals corresponding to the Nb-O bonds ($S_{E+}$-$S_0$) are mostly absent in the spectral difference between bipolar pulses ($S_{E+}$-$S_{E-}$). These experiments confirm the spatially uniform nature of the spectral differences and show that Joule heating and electric field effects are revealed by different species within the cell.

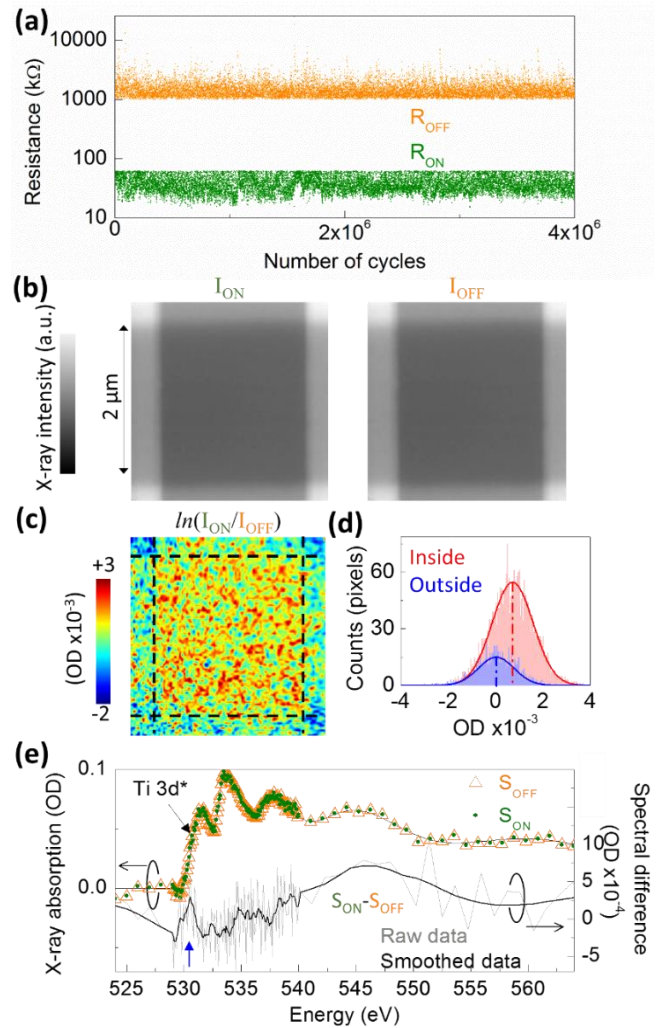

**Figure 3: Resistance switching and material changes.** *(a) Resistance switching of the cell 4 million times between ON and OFF resistance states. (b)-(c) X-ray intensity maps of the ON and OFF states of the cell ($I_{ON}$ and $I_{OFF}$) along with their logarithmic ratio (in the units of optical density, or OD) obtained at an x-ray energy of 530.6 eV. Dashed lines are approximate boundaries of the electrodes. (d) Histograms of the data in (c) inside and outside the crosspoint region. Solid lines are Gaussian fits. Dashed vertical lines indicate the mean of the distribution. Inside the crosspoint, mean (M) = 0.724, standard deviation (S) = 0.721 and standard error of the mean ($M_{err}$) = 0.014, all in units of 10$^{-3}$ OD. Outside the crosspoint, M = 0.017, S = 0.723 and $M_{err}$ = 0.014, all in units of 10$^{-3}$ OD. (e) O K-edge spectra of the region within the crosspoint in the ON ($S_{ON}$) and OFF states ($S_{OFF}$). Solid curve is a fit to the data. Also shown is their spectral difference. Blue arrow points to the sharp feature at ~530.6 eV.*

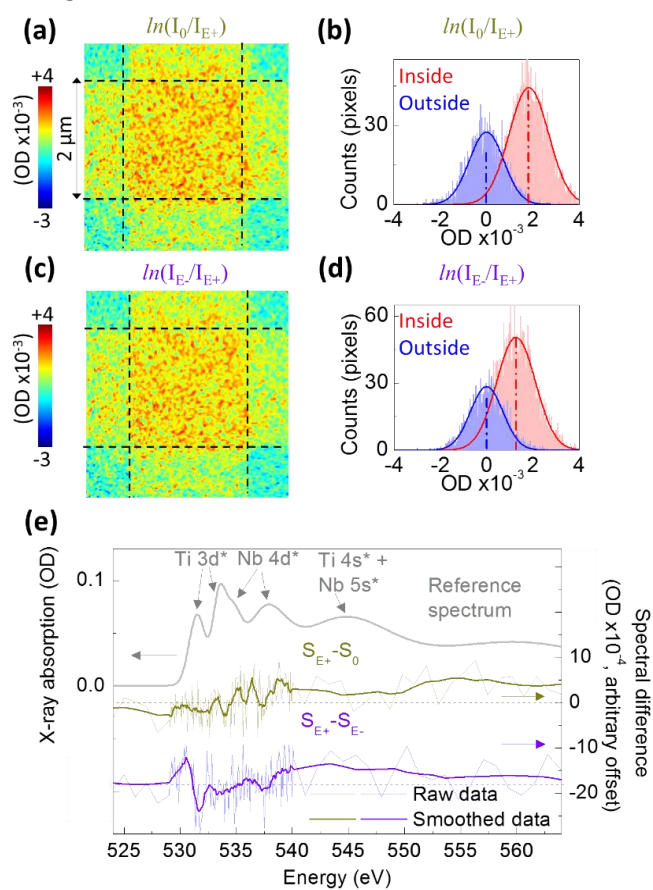

**Figure 4: Joule-heating and electric field effects.** *(a) Logarithmic ratio of intensity maps obtained during application of a voltage pulse and during no application of a voltage pulse at an x-ray energy of 534 eV. (b) Histograms of the data in (a) inside and outside the crosspoint region. (c) Logarithmic ratio of intensity maps obtained during application of voltage pulses of opposite polarity at an x-ray energy of 530 eV. (d) Histograms of the data in (c) inside and outside the crosspoint region. In (b) and (d), solid lines are Gaussian fits, and dashed vertical lines indicate the mean of the distribution. Detailed statistical analyses on this data are shown in Figure S6. (c) Spectral difference within the crosspoint region for the measurements corresponding to (a) and (b). Dashed horizontal lines are zero difference corresponding to the color coded spectral differences. A reference spectrum (from Figure 1b) is included for easy orientation and comparison.*

## Conclusions

In conclusion, we built memristor cells using a trilayer material stack consisting of titanium-niobium-oxides. The cells exhibited favorable performance parameters of low-power switching, low-power reading and high endurance. In order to understand the operational mechanism of the cells, we utilized an *in-operando* time-multiplexed system that could isolate very low signals. This revealed spatially uniform/non-localized material changes caused by switching, in contrast to the frequently observed localized conduction channel formation in simpler material stacks that operate at relatively higher powers. Through O K-edge spectral shifts, we studied the response of different materials in the cell to non-volatile storage and writing of information. Further, the sub-pJ storage energy of the micrometer-sized cell shows promise to scale with area for smaller devices (down to nanometers). These results highlight a novel class of devices, while also stressing the importance of developing non-destructive *in-operando* techniques that can accurately capture the physical mechanisms of the devices.

## Acknowledgements

Synchrotron measurements were performed at the Advanced Light Source (ALS), beamlines 5.3.2.2 and 11.0.2, at Lawrence Berkeley National Laboratory, Berkeley, CA, USA. The ALS is supported by the Director, Office of Science, Office of Basic Energy Sciences, of the U.S. Department of Energy under Contract No. DE-AC02-05CH11231. Work was performed in part at the Stanford Nanofabrication Facility which is supported by National Science Foundation through the NNIN under Grant ECS-9731293.